\documentclass[final,5p,times,twocolumn,numbers]{elsarticle}

\usepackage{amssymb}
\usepackage{lipsum}
\usepackage{amsmath}
\usepackage{graphicx}
\usepackage{caption}
\usepackage{subcaption}
\usepackage{float}
\usepackage{color}
\usepackage{booktabs}

\journal{Nuclear Physics B}

\begin{document}

\begin{frontmatter}

\title{Ion-Optical Tuning of the Large Acceptance Spectrometer \\for Improved Angular Resolution and Acceptance}

\author[first,second]{Taichi Miyagawa\corref{cor1}}\ead{miyatai@rcnp.osaka-u.ac.jp}\cortext[cor1]{Corresponding author}
\author[first,second]{Junki Tanaka}
\author[first]{Shinsuke Ota}
\author[third]{Masanori Dozono}
\author[first]{Nobuyuki Kobayashi}
\author[first]{Lakmin Wickremasinghe}
\author[first,second]{Fumiya Furukawa}
\author[third]{Daichi Ishii}
\author[first,second]{Sho Nishioka}
\author[third]{Kota Takahashi}
\author[third]{Erika Ukai}

\affiliation[first]{organization={Research Center for Nuclear Physics, The University of Osaka},
            addressline={10-1 Mihogaoka},
            city={Ibaraki},
            postcode={567-0047},
            state={Osaka},
            country={Japan}}

\affiliation[second]{organization={Department of Physics, The University of Osaka},
            addressline={1-1 Machikaneyama-cho},
            city={Toyonaka},
            postcode={560-0043},
            state={Osaka},
            country={Japan}}
            
\affiliation[third]{organization={Department of Physics, Kyoto University},
            addressline={Kitashirakawa Oiwake-cho},
            city={Kyoto},
            postcode={606-8502},
            state={Kyoto},
            country={Japan}}

\begin{abstract}
The trade-off between angular resolution and acceptance in scattering-angle measurements with a magnetic spectrometer is quantitatively evaluated for the Large Acceptance Spectrometer (LAS). The dependence on the multipole magnet field strength is investigated.
Third-order transfer matrices were calculated with GICOSY, and particle transport was simulated with MOCADI. The vertical angular resolution is defined as the standard deviation between reconstructed and true angles, while the acceptance is determined from the transport efficiency within an elliptical gate in target angle space.
The resolution improves with increasing field strength, reaching $\sigma_b \sim 5.5$ mrad at +20\%, consistent with $5.43 \pm 0.20$ mrad. In contrast, stronger fields reduce the vertical acceptance and solid angle. These results demonstrate a trade-off between resolution and acceptance. Enhanced vertical focusing shifts the focal condition away from the nominal focal plane, enabling high-precision reconstruction.
\end{abstract}


\end{frontmatter}




\section{Introduction}
\label{introduction}
\vspace{-5pt}

\begin{figure}[h!]
  \centering
  \includegraphics[width=1.0\linewidth]{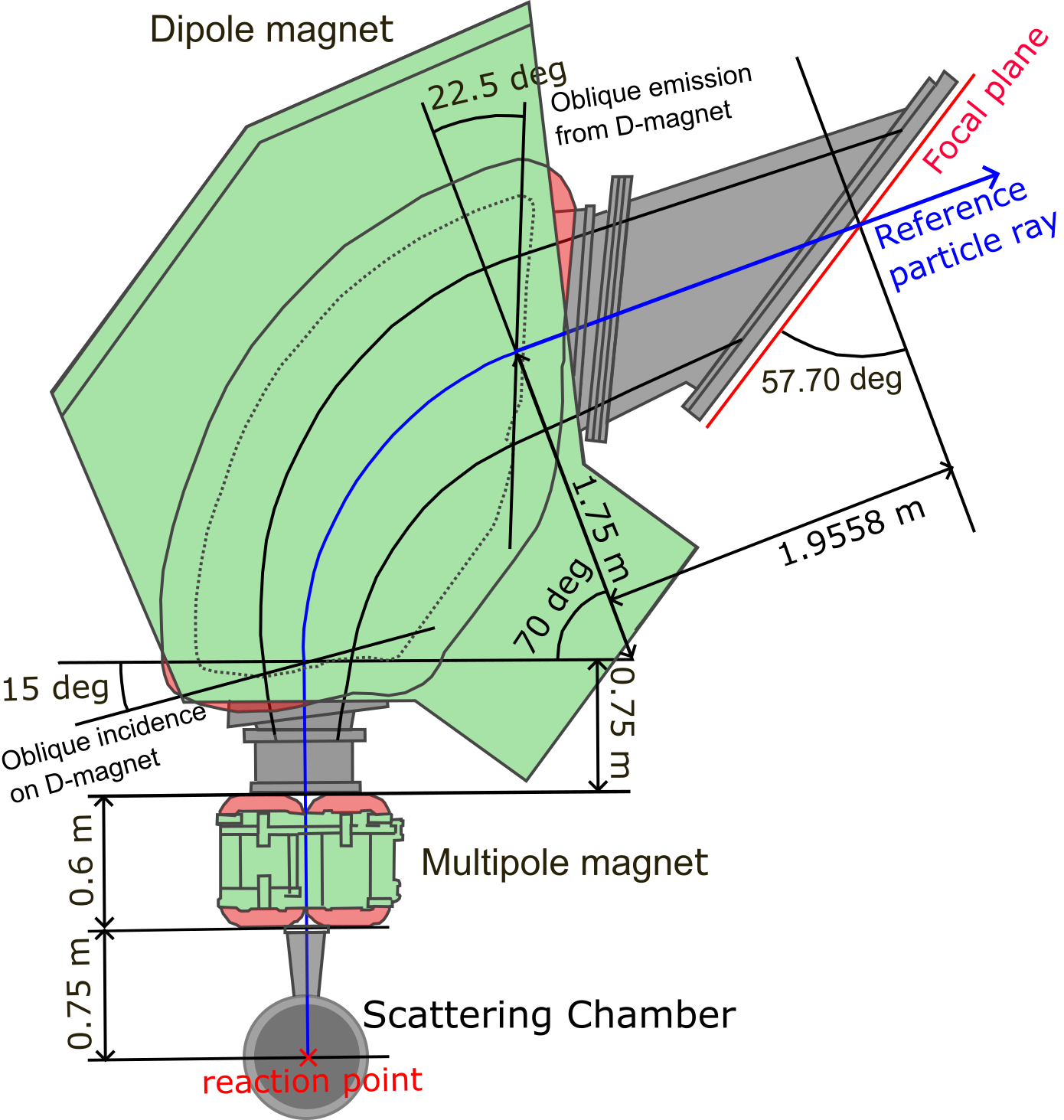}\vspace{-5pt}
  \caption{Schematic view of the geometrical configuration of the Large Acceptance Spectrometer (LAS).}
  \label{fig:LAS}
\end{figure}

Measurement of scattering angles is essential for determining momentum vectors of reaction products and angular distributions of differential cross sections. Magnetic spectrometers are widely used for such measurements. The performance of scattering-angle measurements is evaluated for the Large Acceptance Spectrometer (LAS; Fig.~\ref{fig:LAS}) installed at the Research Center for Nuclear Physics (RCNP), the University of Osaka.
The LAS focal plane has two Vertical Drift Chambers (VDCs) and two plastic scintillator layers. The VDCs measure positions to reconstruct trajectories, while the scintillators provide time-of-flight and energy-loss information. Main characteristics of LAS are summarized in Table~\ref{tab:LAS}.

\begin{table}[htbp]
\centering

\renewcommand{\arraystretch}{1.0} 
\setlength{\tabcolsep}{6pt}        

\caption{Ion-optical and geometrical properties of LAS \cite{Matsuoka1990}.}\vspace{-5pt}
\label{tab:LAS}
\begin{tabular*}{\columnwidth}{@{\extracolsep{\fill}}llll}
\toprule
Resolving Power     & 5,000   & Bend. Radius        & 1.75 m \\
Bending Angle       & 70 deg  & Bend. Power         & 3.22 Tm \\
Dispersion          & 2 m     & Solid Angle         & $\sim 20$ msr \\
Horizontal          & $\pm 60$ mrad & Vertical      & $\pm 100$ mrad \\
Moment. Accept. & 30\%    & Angle range & $0$--$130$ deg \\
\bottomrule
\end{tabular*}
\end{table}

In LAS, scattering angles at the reaction point (subscript $\mathrm{tgt}$) are reconstructed from position and angle information at the focal plane (subscript $\mathrm{fp}$). The vertical scattering angle $b_{\mathrm{tgt}}$ is primarily determined from the vertical position $y_{\mathrm{fp}}$, while the horizontal angle is mainly obtained from the horizontal angle at the focal plane.
In the standard multipole magnet setting, the vertical image $y_{\mathrm{fp}}$ is focused at the focal plane for particles near the reference momentum (see Fig.~\ref{fig:X-Y}(a)). However, strong focusing in $y_{\mathrm{fp}}$ degrades the resolution of the reconstructed vertical scattering angle, whereas a broader $y_{\mathrm{fp}}$ distribution improves the angular resolution.

In this work, the multipole magnet setting is adjusted to intentionally broaden the $y$ distribution at the focal plane. Such conditions, however, reduce the angular acceptance, leading to a trade-off between resolution and acceptance. Therefore, the dependence of angular resolution and acceptance on the multipole magnet field strength is quantitatively evaluated using ion-optical simulations.
\vspace{-5pt}
\section{Properties of Large Acceptance Spectrometer}
\vspace{-5pt}

The Large Acceptance Spectrometer (LAS) consists of a multipole magnet and a dipole magnet (Fig.~\ref{fig:LAS}). The dipole magnet bends particle trajectories in the horizontal plane under a uniform magnetic field, producing momentum-dependent dispersion. As a result, particles with different momenta are spatially separated at the focal plane. Under appropriate geometrical conditions, particles with the same momentum are focused to the same position. This condition is described by Barber’s theorem~\cite{barber}. 
The inclination of the entrance and exit boundaries of the magnetic field modifies particle trajectories, introducing focusing or defocusing effects. These effects contribute not only to the horizontal but also to the vertical focusing properties, and thus play a key role in determining the overall ion-optical performance of the spectrometer.

The multipole magnet employs a Collins-type field configuration, including quadrupole, sextupole, and octupole components. These components control the vertical focusing properties and correct higher-order aberrations that become significant for large solid-angle acceptance. The quadrupole component provides first-order focusing, while the sextupole and octupole components compensate higher-order aberrations.
The magnetic field of the multipole magnet has been reported at $I = 500~\mathrm{A}$~\cite{Matsuoka1992}. The corresponding field gradients for the quadrupole, sextupole, and octupole components are given by Eqs.~(\ref{eq:G1})--(\ref{eq:G3}):
\vspace{-10pt}
\begin{align}
G_1 &= -6.638 \times 10^{-3} \ [\mathrm{T/mm}]\label{eq:G1} \\
G_2 &= 4.889 \times 10^{-6} \ [\mathrm{T/mm^2}]\label{eq:G2} \\
G_3 &= 2.367 \times 10^{-8} \ [\mathrm{T/mm^3}]\label{eq:G3}
\end{align}

The field gradient is defined as
\begin{equation}
G_n = \frac{B_n}{R^{n}}.
\label{eq:gradient}
\end{equation}

The relationship between the applied current $I$ and the magnetic field $|B|$ on the median plane is given by~\cite{Matsuoka1992}:
\begin{align}
|B| &= 0.0151\, I + 0.0312\  (I<350)\\
|B| &= -2.52\times10^{-5}\, I^{2} + 0.0315\, I - 2.614\  (I\geqq350)
\end{align}

Since the ratios of the multipole components remain constant with $I$, the relation
$G_n(I) : G_n(I') = B(I) : B(I')$
holds. Therefore, using the known coefficients at $I = 500~\mathrm{A}$, the field gradient at an arbitrary current $I$ can be expressed as
\begin{equation}
G_n(I) = \frac{B(I)}{B(500)} \, G_n(500).
\label{eq:MQ}
\end{equation}

These relations are used to reproduce the multipole magnet field in the present simulation.
\vspace{-5pt}
\section{Ion-Optical Simulation}
\vspace{-5pt}

Ion-optical calculations were performed using the code GICOSY~\cite{Wollnik_GICOSY}, and Monte Carlo particle transport simulations were carried out with MOCADI~\cite{Iwasa1997}. Third-order transfer matrices of LAS were first calculated with GICOSY and then used as input for the particle transport simulations in MOCADI.
A right-handed coordinate system was adopted with the beam axis defined as the $z$ axis. A particle is described by $(x, a, y, b, \delta)$ at a given $z$, where $x$ and $y$ denote the horizontal and vertical positions, $a$ and $b$ the corresponding angles, and $\delta$ the relative momentum deviation from the reference momentum. Subscripts $\mathrm{tgt}$ and $\mathrm{fp}$ denote quantities at the target and the focal plane, respectively.
$\alpha$ particles ($A = 4$, $Z = 2$) with an energy of $62.8~\mathrm{MeV}$ were used as representative incident particles corresponding to the reference magnetic rigidity. The present results are determined by the magnetic rigidity and are therefore not specific to the particle species. The reference magnetic rigidity was set to $B\rho = 1.146~\mathrm{Tm}$, corresponding to the conditions of previous experiments. The initial position distribution was defined as a circular beam spot with radii of $1~\mathrm{mm}$ in both horizontal and vertical directions at the target. The initial angular distributions were assumed to be uniform within $\pm 80~\mathrm{mrad}$ in the horizontal direction and $\pm 130~\mathrm{mrad}$ in the vertical direction. These ranges cover the geometrical acceptance of LAS. The initial momentum deviation was also uniformly distributed within $\pm 20\%$.

\begin{figure}[h]
  \centering
  \includegraphics[width=1.0\linewidth]{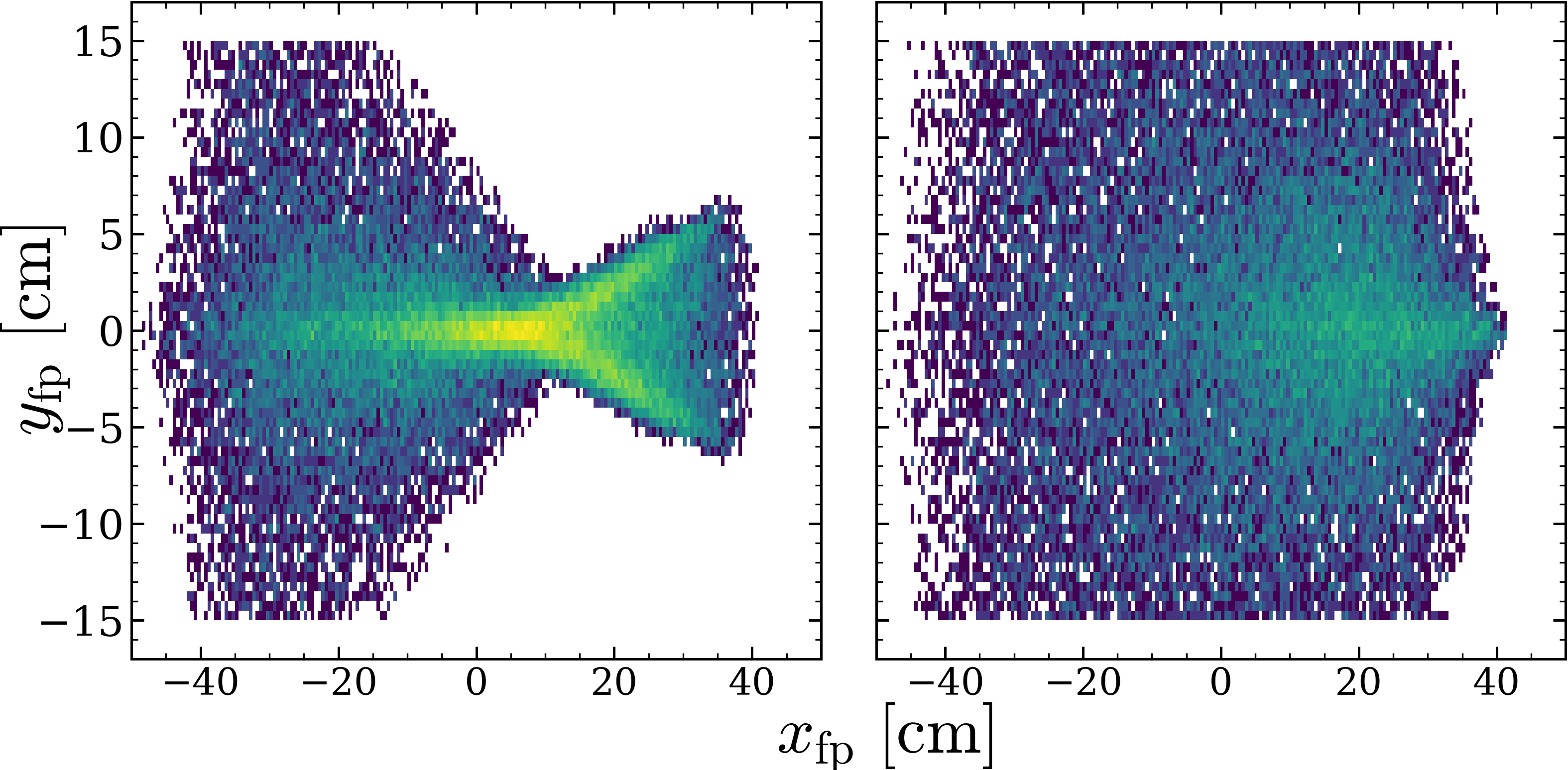}\vspace{-5pt}
  \caption{Simulated $X_{\mathrm{fp}}$–$Y_{\mathrm{fp}}$ distributions at the LAS focal plane for (a) the standard setting and (b) a +20\% multipole magnet field.}
  \label{fig:X-Y}
\end{figure}

The magnetic field strength of the multipole magnet was varied within $\pm 20\%$ around the standard setting, and simulations were performed for each condition. The standard setting corresponds to the condition where particles with the reference magnetic rigidity ($X_{\mathrm{fp}} \sim 0$) are focused in the vertical direction at the exit of LAS, as shown in Fig.~\ref{fig:X-Y}(a).
The particle distributions at the LAS exit were obtained for each magnetic field condition. For the case of a $+20\%$ field strength, the vertical focal position shifts toward positive $X_{\mathrm{fp}}$, as shown in Fig.~\ref{fig:X-Y}(b). This behavior is consistent with previous experimental observations~\cite{Tanaka2021,Matsumura2026}.

\vspace{-5pt}
\section{Reconstructed scattering angular resolution}
\label{sec:angular_resolution}
\vspace{-5pt}

The reconstruction resolution of the vertical scattering angle $b_{\mathrm{tgt}}$ at the reaction point is evaluated. In the forward simulation, particle positions and angles at the focal plane, $(x_{\mathrm{fp}}, a_{\mathrm{fp}}, y_{\mathrm{fp}}, b_{\mathrm{fp}})$, are obtained using the transport matrix. These quantities are measured with finite detector resolutions and used to reconstruct the scattering angle at the target.

\begin{figure}[htbp]
  \centering
  \includegraphics[width=0.9\linewidth]{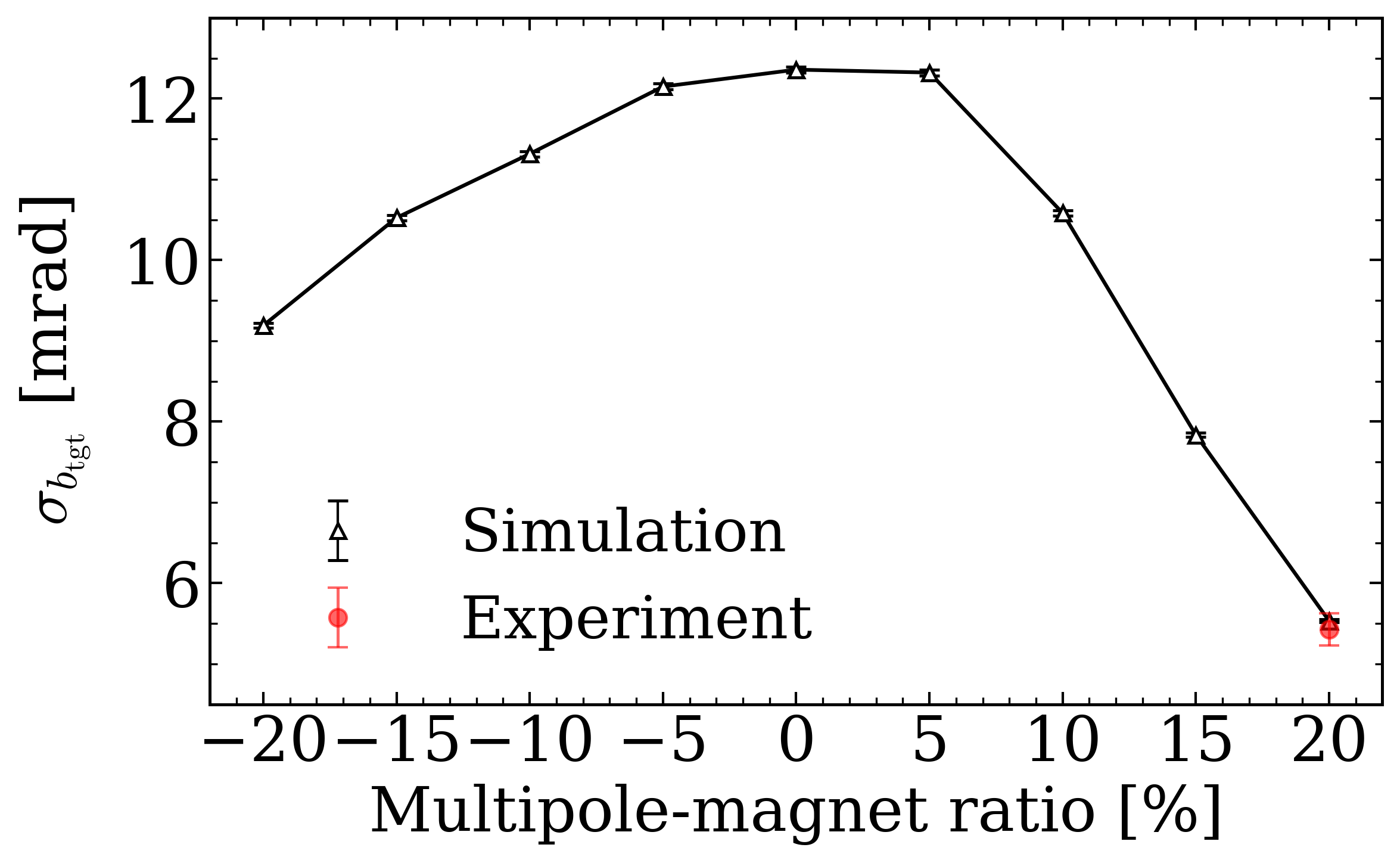}\vspace{-5pt}
  \caption{Dependence of the vertical angular resolution $\sigma_{b_{\mathrm{tgt}}}$ on the multipole magnet field strength. Black triangles represent simulations, and red points indicate experimental data~\cite{Matsumura2026}.}
  \label{fig:Btgt_resolution}
\end{figure}

To simulate this measurement process, Gaussian fluctuations corresponding to the detector resolutions are applied to each focal-plane variable, generating pseudo-measured values. The inverse transport matrix is then applied to reconstruct the scattering angle $b_{\mathrm{rec}}$ at the target.
Because the relation includes higher-order nonlinearities, the inverse transport is expressed as a polynomial expansion up to third order. The reconstructed vertical scattering angle is written as
\begin{equation}
b_{\mathrm{rec}} = \sum_i c_i x^{p_i} a^{q_i} y^{r_i} b^{s_i},
\end{equation}
where $c_i$ are coefficients and $p_i, q_i, r_i, s_i$ are the exponents of each variable satisfying $p_i + q_i + r_i + s_i \leq 3$. The coefficients are determined by least-squares fitting using a design matrix constructed from the polynomial basis up to third order.
The detector resolutions are assumed to be $\sigma_{x_{\mathrm{fp}}}=\sigma_{y_{\mathrm{fp}}}=200~\mu\mathrm{m}$ and $\sigma_{a_{\mathrm{fp}}}=\sigma_{b_{\mathrm{fp}}}=3~\mathrm{mrad}$, corresponding to typical VDC performance. The angular resolution is defined as the standard deviation of the residual $\Delta b = b_{\mathrm{rec}} - b_{\mathrm{tgt}}$.
The dependence of $\sigma_{b_{\mathrm{tgt}}}$ on the multipole magnet field strength is shown in Fig.~\ref{fig:Btgt_resolution}. The resolution improves as the field strength increases, reaching $\sigma_{b_{\mathrm{tgt}}} \sim 5.5~\mathrm{mrad}$ at $+20\%$. In contrast, the standard setting yields $\sigma_{b_{\mathrm{tgt}}} \sim 12~\mathrm{mrad}$, indicating that strong focusing in $y_{\mathrm{fp}}$ degrades the angular resolution.

\vspace{-5pt}

At $+20\%$, the experimental value $\sigma_b = 5.43 \pm 0.20~\mathrm{mrad}$~\cite{Matsumura2026} agrees with the simulated value $\sigma_b = 5.53 \pm 0.02~\mathrm{mrad}$ within uncertainties. This agreement confirms the validity of the reconstruction procedure, including the inverse transport and detector resolution effects.
\vspace{-5pt}
\section{Evaluation of Angular Acceptance}
\label{sec:angular_acceptance}
\vspace{-5pt}

The magnetic field strength of the multipole magnet primarily controls vertical focusing and strongly affects the angular acceptance. In this section, the angular acceptance is quantitatively evaluated based on simulations.
Events with initial angles $(a_{\mathrm{tgt}}, b_{\mathrm{tgt}})$ are generated at the target and transported through the spectrometer. Only events that reach the focal plane are regarded as detected. These events are traced back to obtain reconstructed angles $(a_{\mathrm{rec}}, b_{\mathrm{rec}})$. Since events outside the acceptance are lost during transport, the reconstructed angular distribution is effectively limited by the angular and momentum acceptance.

To evaluate the angular acceptance, an elliptical region is introduced in the target angular space:
\begin{equation}
\frac{a_{\mathrm{tgt}}^2}{a_{\mathrm{acc}}^2} + \frac{b_{\mathrm{tgt}}^2}{b_{\mathrm{acc}}^2} \leq 1.
\label{eq:gate}
\end{equation}
Here, $a_{\mathrm{acc}}$ and $b_{\mathrm{acc}}$ represent the horizontal and vertical acceptance limits, respectively. The elliptical shape allows independent control of horizontal and vertical limits. The transport efficiency $\varepsilon$ is defined as the fraction of generated events within this region that reach the focal plane.
The parameters $a_{\mathrm{acc}}$ and $b_{\mathrm{acc}}$ are scanned over $10$--$99~\mathrm{mrad}$ and $10$--$149~\mathrm{mrad}$, respectively, with a step size of $1~\mathrm{mrad}$. Among all combinations, those satisfying the beam transmission $\varepsilon = 1$ are selected, and the set that maximizes the solid angle
\begin{equation}
\Omega = \pi a_{\mathrm{acc}} b_{\mathrm{acc}}
\end{equation}
is defined as the angular acceptance for each magnetic field condition. The uncertainty due to the step size is estimated to be $\pm 0.5~\mathrm{mrad}$. This definition ensures that the acceptance is free from transport losses.
This procedure uniquely determines, for each magnetic field setting, the largest angular region without transport loss.

\begin{figure}[htbp]
  \centering
  \includegraphics[width=1.0\linewidth]{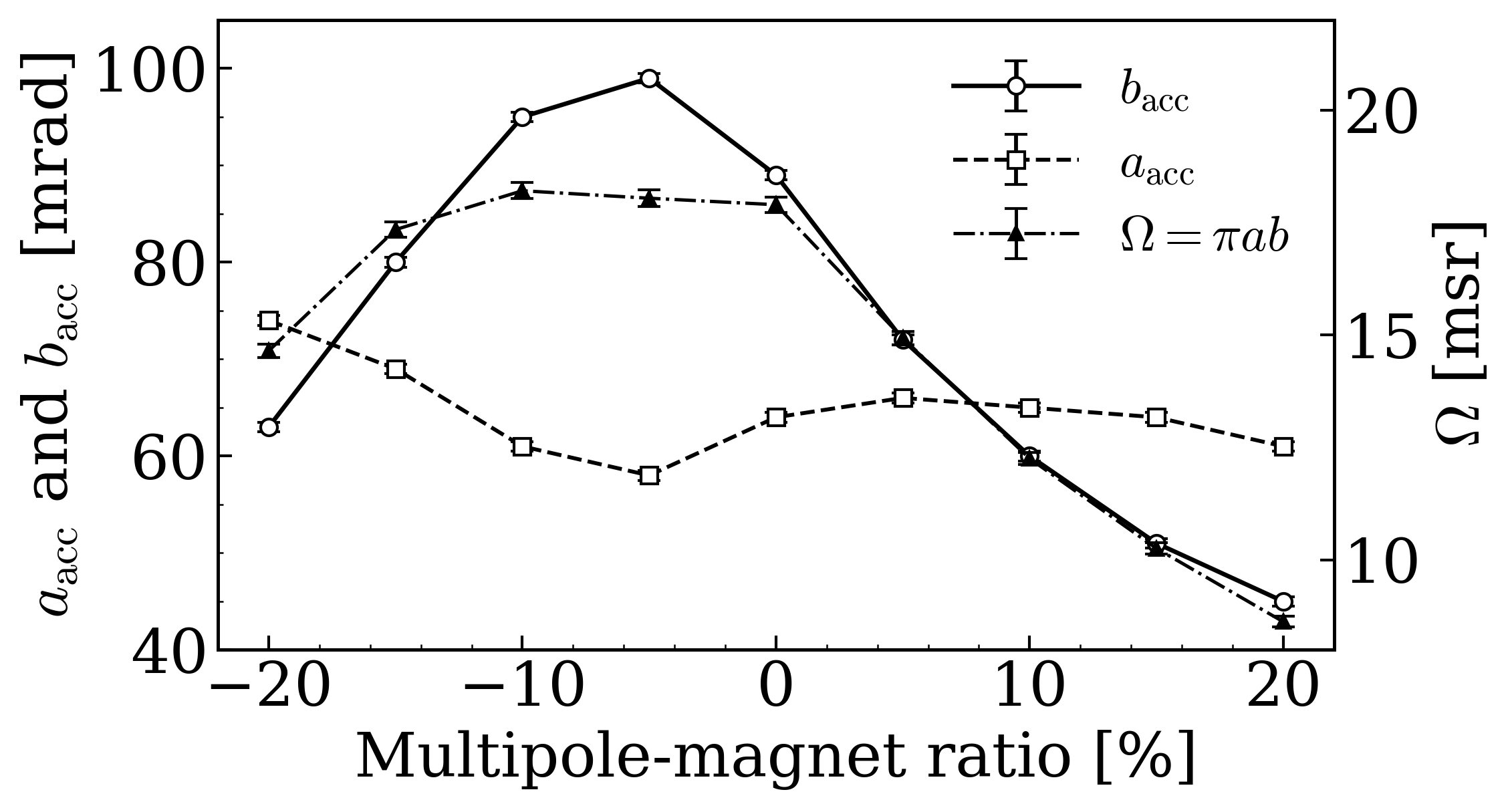}\vspace{-5pt}
  \caption{Angular acceptance and corresponding solid angle as functions of the multipole magnet field strength.}
  \label{fig:angular_acceptance}
\end{figure}

Figure~\ref{fig:angular_acceptance} shows the dependence of the angular acceptance and corresponding solid angle on the multipole magnet field strength. The vertical acceptance $b_{\mathrm{tgt}}$ increases as the field strength decreases and reaches a maximum around $-5\%$, while the horizontal acceptance $a_{\mathrm{tgt}}$ shows only a weak dependence and remains approximately $60$--$70~\mathrm{mrad}$.
As a result, the solid angle $\Omega$ is mainly governed by the variation of $b_{\mathrm{tgt}}$, with a maximum in the range of $-10\%$ to $-5\%$. This shows that the angular acceptance is primarily determined by the vertical focusing properties.
\vspace{-10pt}
\section{Conclusion}
\label{sec:conclusion}
\vspace{-5pt}

The effects of the multipole magnet field strength on the vertical angular resolution and angular acceptance in the Large Acceptance Spectrometer (LAS) have been evaluated using ion-optical calculations with GICOSY and particle transport simulations with MOCADI.
The vertical angular resolution improves with increasing field strength, reaching an optimal value of $\sigma_b \sim 5.5~\mathrm{mrad}$ at $+20\%$, in good agreement with the experimental value of $5.43 \pm 0.20~\mathrm{mrad}$. In contrast, stronger fields reduce the vertical angular acceptance and the corresponding solid angle. These results demonstrate a trade-off between angular resolution and acceptance.
A stronger multipole magnet setting is advantageous for achieving high angular resolution, whereas the standard setting is more suitable for maximizing angular acceptance.
This provides practical guidance for selecting optimal optical conditions depending on experimental requirements.
\section*{Acknowledgements}
The authors thank the RCNP supporting staff for their technical assistance. This work was supported by JSPS KAKENHI Grant Number 25H00640. J.T. acknowledges support from the Yamada Science Foundation.


\vspace{-10pt}
\begin{thebibliography}{9}
\bibitem{Matsuoka1990}
N.~Matsuoka,
RCNP Annual Report 1990, Research Center for Nuclear Physics, Osaka University (1990).
\bibitem{Matsuoka1992}
N.~Matsuoka,
RCNP Annual Report 1992, Research Center for Nuclear Physics, Osaka University (1992).
\bibitem{barber}
N.F.~Barber,
Proc.\ Leeds.\ Phil.\ Soc.\ \textbf{2}, 427(1933).
\bibitem{Wollnik_GICOSY}
H.~Wollnik,
GSI Helmholtzzentrum f\"ur Schwerionenforschung, Darmstadt, internal report.
\bibitem{Iwasa1997}
N.~Iwasa, T.~Ohnishi, and H.~Geissel,
Nucl.\ Instrum.\ Methods Phys.\ Res.\ B \textbf{126}, 284--290 (1997).
\bibitem{Tanaka2021}
J.~Tanaka, Z.H.~Yang, S.~Typel et al.,
Science \textbf{371}, 260--264 (2021).
\bibitem{Matsumura2026}
R.~Matsumura, J.~Tanaka, K.~Yoshida, et al.,
Prog. Theor. Exp. Phys., 043D01 (2026).
\end{thebibliography}
\end{document}